\documentclass[a4paper, twocolumn, 11pt, accepted=2022-04-18]{quantumarticle}
\pdfoutput=1
\usepackage[utf8]{inputenc}
\usepackage[english]{babel}
\usepackage[T1]{fontenc}
\usepackage{amsmath}
\usepackage{hyperref}

\usepackage{tikz}

\usepackage[version=3]{mhchem} 
\usepackage{algorithm}
\usepackage{algorithmic}
\usepackage{physics}
\usepackage[numbers,sort&compress]{natbib}

\usepackage[showframe]{geometry}
\usepackage{graphicx}
\newcommand{\change}[1]{{#1}} 
\newcommand{\changed}[1]{{#1}} 

\begin{document}

\title{Unitary Selective Coupled-Cluster Method}

\author{Dmitry A. Fedorov}
\affiliation
{Computational Science Division, Argonne National Laboratory, 9700 S. Cass Ave., Lemont, IL 60439, USA}
\affiliation{Oak Ridge Associated Universities, Oak Ridge, TN 37830}
\orcid{0000-0003-1659-8580}
\author{Yuri Alexeev}
\affiliation
{Computational Science Division, Argonne National Laboratory, 9700 S. Cass Ave., Lemont, IL 60439, USA}
\orcid{0000-0001-5066-2254}
\author{Stephen K. Gray}
\affiliation{Center for Nanoscale Materials, Argonne National Laboratory, 9700 S. Cass Ave, Lemont, IL 60439, USA}
\orcid{0000-0002-5188-354X}
\author{Matthew J.Otten}
\email{mjotten@hrl.com}
\affiliation{HRL Laboratories, LLC, 3011 Malibu Canyon Road,
Malibu, CA 90265}
\orcid{0000-0002-6522-5820}

\maketitle

\begin{abstract}
  Simulating molecules using the Variational Quantum Eigensolver method is one of the promising applications for NISQ-era quantum computers. Designing an efficient ansatz to represent the electronic wave function is crucial in such simulations. Standard unitary coupled-cluster with singles and doubles (UCCSD) ansatz tends to have a large number of insignificant terms that do not lower the energy of the system. In this work, we present a unitary selective coupled-cluster method, a way to construct a unitary coupled-cluster ansatz iteratively using a selection procedure with excitations up to fourth order. This approach uses the electronic Hamiltonian matrix elements and the amplitudes for excitations already present in the ansatz to find the important excitations of higher order and to add them to the ansatz. The important feature of the method is that it systematically reduces the energy error with increasing ansatz size for a set of test molecules. \change{The main advantage of the proposed method is that the effort to increase the ansatz does not require any additional measurements on a quantum computer.}
\end{abstract}

\section{Introduction}
One of the promising and natural applications of quantum computers is the simulation of quantum systems. By representing a state of a quantum system on a quantum computer, one can, in theory, avoid the exponential scaling faced on classical computers when manipulating the state of a quantum system. For example, it has been proven that the quantum phase estimation (QPE) algorithm can find eigenvalues of unitary operators exponentially faster than a classical computer can if a trial state with nonzero overlap with the solution can be prepared on a quantum computer.\cite{kitaev_QPE_1995, QPE_Abrams_1997, QPE_Lloyd} Since the time propagator of a quantum system is a unitary operator, QPE seems like a good candidate for solving many problems from physics and chemistry on quantum computers. \change{Although the QPE algorithm for chemical systems has been demonstrated on a simulator,\cite{qpe_chemistry} such simulations on quantum hardware are far beyond the capabilities of current, noisy intermediate-scale quantum (NISQ) computers.\cite{preskill2018quantum, elfving2020will}} The Variational Quantum Eigensolver (VQE)\cite{Peruzzo2014_VQE} is a hybrid quantum-classical algorithm to find eigenvalues of molecular Hamiltonians that requires shallower circuits than QPE. The shallower circuits are achieved at the expense of a larger number of measurements and by offloading a part of the computational workload onto a classical computer. In VQE simulations, it is important to choose an efficient variational ansatz that (a) is expressible enough so that it can represent a good solution to the problem, (b) is not too expressible so that the solution space is limited to relevant parts of Hilbert space, (c) does not have too many parameters to optimize, and (d) can be represented by short circuits.\cite{fedorov2021vqe}

Conceptually, one can approach the problem of designing a variational ansatz in two ways. The first, so-called chemistry-inspired type of ansatzes, is based on the unitary coupled-cluster (UCC) method. The unitary coupled-cluster with singles and doubles (UCCSD)\cite{Taube2006, Evangelista2019, Cooper2010, Harsha2018, Pal1984use} ansatz was used to run the first VQE simulation.\change{\cite{Peruzzo2014_VQE, McClean_2016,romero2018, anand2021quantum}} This ansatz is inspired by the commonly used classical coupled-cluster method. Essentially, each parameter in a UCCSD ansatz parameterizes a coupled-cluster amplitude for each fermionic excitation from a reference state, either a single or a double excitation. This ansatz has been successfully applied to compute accurate electronic energies of small molecules. Because of the physically motivated structure, it is easy to optimize such ansatzes because they generate states that correspond to realistic electronic wave functions. However, such ansatzes for a large molecule can have many redundant, unimportant excitations and, therefore, a large number of parameters to optimize and excessively long circuits. This problem occurs because the only variables used when generating an ansatz are the number of orbitals and electrons, without taking into account the information about the specific chemical system, such as geometry or point group symmetry.

The second group of ansatzes is called ``hardware-efficient'' Such ansatzes contain sequences of parameterized single- and two-qubit gates that can be easily implemented on the quantum computer at hand.\cite{kandala_he_ansatz, kandala_2019_he_on_hardware, Barkoutsos_2018, Ganzhorn2019, gard_sym_preserv} No information about the physics of the molecule being studied is used. Hardware-efficient ansatzes are very expressible. Furthermore, the circuits to implement them are shallower compared with UCCSD ones and contain a smaller number of error-prone two-qubit quantum gates. However, such ansatzes can also have too many parameters to optimize and suffer from the ``barren plateaus''\cite{barren_plateaus, wang2021noiseinduced, Cerezo2021} problem. In addition, because of the lack of physically motivated structure, extra work needs to be done to enforce physical symmetries, such as antisymmetry of the wave function and particle number.\cite{gard_sym_preserv}

A number of recent studies have focused on designing ansatzes iteratively by adding elements into the ansatz based on some system-specific criterion compared with a general-purpose ansatz such as UCCSD~\cite{Grimsley2019, Tang_qubit_adapt, qcc, it_QCC, lang2020unitary, humble_benchmarking, sim2020adaptive, Kottmann2021}. The first iterative algorithm, ADAPT-VQE~\cite{Grimsley2019}, adds fermionic operators to the ansatz based on the gradients of energy with respect to the variational parameters associated with these operators. The choice of an operator to add at each step is performed by computing the energy gradients with respect to variational parameters. This screening procedure allows one to pick operators that will contribute the most to the lowering of energy. The criterion to choose the operators is borrowed from the anti-Hermitian contracted Schr\"{o}dinger equation  methodology.\cite{acse_Mazziotti2004, exact_2body_exp_mazziotti2020, Mazziotti2006, mazziotti2007, Mukherjee2001} 

In a variation of the ADAPT-VQE method, called qubit-ADAPT-VQE~\cite{Tang_qubit_adapt}, the fermionic operators are broken down into Pauli strings and used as building blocks for constructing an ansatz. As a result, the circuits have fewer two-qubit gates, which makes this ansatz more NISQ-friendly. However, the number of variational parameters is significantly larger than in ADAPT-VQE, which makes classical optimization harder and increases the number of ADAPT iterations. \change{Another method that uses operators constructed in the qubit space is} the qubit coupled-cluster (QCC)\cite{qcc} method and its iterative version\cite{it_QCC}. Similar to ADAPT-VQE, the decision on which operator to add to the ansatz is based on the gradients of energy with respect to variational parameters corresponding to these operators.

An alternative way to increase the efficiency of a fermionic operator-based ansatz is to use the prescreening procedure based on the $t_2$ amplitudes, coefficients computed by using M\o{}ller-Plesset second-order perturbation (MP2) theory~\cite{Head-Gordon1988, Moller1934} corresponding to double-excitation fermionic operators. \change{This approach has been demonstrated to significantly reduce the quantum gate count for certain molecular systems.\cite{romero2018}}

\section{VQE-UCC Framework}
We begin with a brief introduction to the VQE method and the UCC ansatz, which is necessary for understanding the rest of this work. A more detailed description can be found in several recent, comprehensive reviews.\change{\cite{quant_comp_chem_revmodphys, quant_chem_chemrev_2019, VQA_rev_2020, bharti2021noisy, tilly2021variational}} The goal is to solve the time-independent Schr\"{o}dinger equation for electrons, which is obtained by applying the Born--Oppenheimer approximation for separation of electronic and nuclear degrees of freedom. This equation is an eigenvalue problem $H_{el}\ket{\psi}=E\ket{\psi}$, where $\ket{\psi}$ is the electronic wave function, $E$ is the energy of the ground electronic state, and $H_{el}$ is an electronic Hamiltonian, which can be written by using the second quantization formalism as
\begin{equation}\label{el_ham}
    H_{el} = \sum_{pq}^{N_{MO}}h_{pq}a_p^{\dagger}a_q+\sum_{pqrs}^{N_{MO}}h_{pqrs}a_p^{\dagger}a_q^{\dagger}a_sa_r .
\end{equation}
In Equation \ref{el_ham}, indices $p$, $q$, $r$,and $s$ denote the molecular spin orbitals, $N_{MO}$ is the total number of molecular spin orbitals, and $h_{pq}$ and $h_{pqrs}$ are the electronic Hamiltonian matrix elements connecting corresponding spin orbitals, which are easily calculated on a classical computer. These electronic Hamiltonian matrix elements are also referred to as one- and two-electron integrals. We will use the two terms interchangeably throughout the paper. Here $a_p^{\dagger}$ and $a_q$ are creation and annihilation operators, respectively, which add or remove an electron from orbitals $p$ and $q$. The second-quantized fermionic Hamiltonian can be mapped onto the qubit space by using one of the fermion-to-qubit mappings. \change{The resulting qubit Hamiltonian is used to measure the energy on a quantum computer. For more details we refer the reader to the literature on the topic.\cite{quant_comp_chem_revmodphys, quant_chem_chemrev_2019, VQA_rev_2020, bharti2021noisy, tilly2021variational}} 

In this study we work with the unitary coupled-cluster (UCC) ansatz, which, as noted above, is a so-called chemistry-inspired ansatz. It is a unitary version of the classical coupled-cluster method, a high-accuracy method in classical quantum chemistry.\cite{cc_purvis_1978,Purvis1982, cc_review_bartlett} Unlike the regular coupled-cluster (CC)  theory, however, UCC can be efficiently implemented on a quantum computer. The wave function is parameterized by using a unitary operator,
\begin{equation}\label{uccsd_exc_op}
 \ket{\psi(\vec{t})}=e^{\hat{T}-\hat{T}^{\dagger}}\ket{\phi},
\end{equation}
where $\hat{T}$ is the cluster excitation operator. Since including all possible excitations is not computationally feasible because of exponential scaling with system size, the CC expansion is truncated, most commonly at the doubles level, which corresponds to the UCCSD ansatz
\begin{equation}\label{uccsd_t_operator}
    \begin{split}
        \hat{T} = \hat{T}_1 + \hat{T}_2 = \sum_{i\in{occ},a\in{virt}}\hat{t}_{i}^{a}+ \sum_{i,j\in{occ},a,b\in{virt}}\hat{t}_{ij}^{ab} = \\ 
        = \sum_{i\in{occ},a\in{virt}}t_{i}^{a}\hat{a}^{\dagger}_a\hat{a}_i+ \sum_{i,j\in{occ},a,b\in{virt}}t_{ij}^{ab}\hat{a}^{\dagger}_a\hat{a}^{\dagger}_b\hat{a}_i\hat{a}_j.
    \end{split}
\end{equation}
where $t_{i}^a$ and $t_{ij}^{ab}$ are the cluster amplitudes, each corresponding to an excitation from occupied orbitals $i$ and $j$ into virtual orbitals $a$ and $b$. \change{Here occupied orbitals are the spin orbitals that are occupied in a reference state,  the Hartree-Fock state. Virtual orbitals are the spin orbitals, to which electrons are excited from the reference state by applying the cluster operators described above. The direct implementation of an exponential operator in Equation~\eqref{uccsd_t_operator}} with many terms that act on all qubits is not possible on quantum hardware. The operator needs to be broken into a sequence of operators acting on a few qubits each. Trotterization using the Suzuki--Trotter expansion\cite{trotterization} is the most common approach to achieve this goal and is defined as
\begin{equation}\label{trotterization}
    e^{A+B} = \lim_{n \to \infty} (e^{A/n}e^{B/n})^n.
\end{equation}
Because of the variational flexibility of the UCC ansatz, the Trotter error remains small even if a single Trotter step ($n=1$) is used.\cite{omalley2016}

\change{In VQE-UCCSD framework, cluster amplitudes $t_{i}^a$ and $t_{ij}^{ab}$} are variationally optimized in a classical loop. Although all generated ansatz elements are inspired by chemistry, with a single parameter that controls the contribution of a particular excitation, the UCC ansatz has a drawback that can make it inefficient for large systems. Specifically, the UCC ansatz depends only on the number of molecular orbitals and electrons; it does not take into account the information about the chemistry of the particular molecule. For example, different molecules will have different sets of important excitations. As a result, the UCC ansatz can contain a large fraction of excitations that are insignificant or are simply zero because of the presence of symmetry. Since implementing each excitation on quantum hardware is expensive and the total number of possible excitations grows rapidly with system size, considerable effort has been expended in designing more efficient UCC ansatzes. The focus of this work is on constructing an efficient UCC ansatz that includes the ``most important" fermionic excitation operators that contribute the most to the correlation energy.

\section{Selected CI}
Within classical algorithms for quantum chemistry there exist many techniques for calculating estimates of the ground state energy, including the coupled-cluster methods (such as CCSD) mentioned above~\cite{Purvis1982}. Another family of algorithms works in the determinant basis, rather than with cluster operators. Full configuration interaction (FCI) finds the coefficients of the combinatorially many determinants of the many-body wave function through an exact diagonalization-like approach~\cite{knowles1984new}. 

Selective configuration interaction algorithms, such as perturbatively selected configuration interaction (CIPSI)~\cite{evangelisti1983convergence}, adaptive sampling configuration interaction (ASCI)~\cite{tubman2016deterministic}, and selected heat-bath configuration interaction (SHCI)~\cite{holmes2016heat,li2018fast}, use an iterative procedure to select only important determinants to include in the wave function, regardless of the order of excitation. 

Although this is still exponentially scaling, it often leads to a dramatic reduction in the number of determinants needed to reach accurate calculations, allowing for FCI-quality energies in systems as large as the highly correlated chromium dimer, correlating 28 electrons in 198 orbitals (a Hilbert space of size $\approx 10^{42})$~\cite{li2020accurate}. SCI methods can also be used to find excited states, through a modified selection criterion~\cite{holmes2017excited,chien2018excited}.

Here, we briefly review the determinant selection step of the SHCI algorithm, since some of its features inspired the unitary selected coupled-cluster algorithm that we discuss below. A more complete description of the SCHI algorithm can be found in Refs.~\cite{holmes2016heat,li2018fast} In the variational step of SHCI, determinants connected to the current variational set of determinants $\mathcal{V}$ (which, at the first iteration, is typically just the Hartree--Fock determinant) are screened for importance via the selection criterion
\begin{equation}\label{shci_criterion}
    \max_{D_i\in \mathcal{V}} |H_{ia} c_i| > \epsilon_1,
\end{equation}
where $H_{ia}$ is the Hamiltonian matrix element connecting determinant $D_a$, which is not currently in the variational set $\mathcal{V}$, to determinant $D_i$, which is currently in the variational set with coefficient $c_i$. Here $\epsilon_1$ is a user-defined cutoff that controls the accuracy of the method; in the limit that $\epsilon_1$ is 0, the method will include all determinants and is equivalent to FCI. Other SCI methods, such as CIPSI~\cite{evangelisti1983convergence}, use a selection criterion based on perturbation theory. The SHCI selection criterion has the intuitive interpretation of being the  ``most important" part of the perturbative correction~\cite{holmes2016heat}. It can also be interpreted, loosely, as including determinants that are strongly connected (having a large matrix element $H_{ia}$) to an important determinant (one that has a large coefficient, $c_i$).

\subsection{Unitary Selective coupled-cluster}
The USCC method we propose is an iterative method to construct a unitary coupled-cluster ansatz, where the excitation operators are added to the ansatz based on the screening criterion:
\change{
\begin{equation} \label{uscc_selection}
|H_{\beta a}t_{\beta}| \geq \epsilon .
\end{equation}

In Equation~\eqref{uscc_selection}, $t_\beta$ are the amplitudes (of any order)} corresponding to the excitation operators already included in the ansatz (and determined by VQE optimization). Similarly to the selected CI method (see Eq.~\eqref{shci_criterion}), $H_{ia}$ are the electronic Hamiltonian matrix elements $h_1$ and $h_2$ connecting the spin orbitals involved in excitations from the reference state defined by amplitudes $t_i$ and $t_a$. This criterion is inspired by the SHCI criterion, replacing the determinantal coefficient $c_i$ with the cluster amplitude $t_i$. A first-order expansion of the UCC operator leads to the rough approximation $c_i \approx t_i$. This can be observed in small systems for which there is a one-to-one mapping between determinants and cluster excitation operators; that is, for a certain determinant there is only one excitation operator that connects it to the Hartree--Fock state. In such a case cluster amplitude is equal to the CI coefficient, for example in the \ce{H2} molecule in the minimal basis set. Another way to look at the unitary selective coupled-cluster selection criterion, Equation~\eqref{uscc_selection}, is that when $t_i \approx 0$, the Trotter step corresponding to that cluster amplitude is nearly identity, meaning that it has little effect on the wave function, although care must be taken because of the periodic nature of the UCC form. 

The ansatz generation starts with the reference Hartree--Fock state, and none of the excitation amplitudes are available. First, all possible single- and double-electronic excitations are generated. The excitations are written in spin-block notation: the occupied orbitals, from which excitations are performed, are labeled $i$, $j$, $k$, and $l$; and the virtual orbitals, to which electrons are excited, are labeled $a$, $b$, $c$, and $d$. For example, single and double excitations are written as $[i,a]$ and $[i,j,a,b]$. The electron integrals between these orbitals are respectively. All $[i, a]$ and $[i, j, a, b]$ excitations, for which the absolute values of corresponding $h_1[i, a]$ and $h_2[i, j, a, b]$ matrix elements are larger than the threshold value for the first iteration $\epsilon_1$, are included in the ansatz. Then, this initial ansatz is used to run the first VQE simulation. All $t_1[i, a]$ and $t_2[i, j, a, b]$ amplitudes are saved for future iterations. Thus, each single (double) excitation has two coefficients associated with it, $h_1$ and $t_1$ ($h_2$ and $t_2$).
    
\begin{figure}[tp]
\centering
    \includegraphics[width=0.48\textwidth, 
                     height = 0.48\textheight,  keepaspectratio]{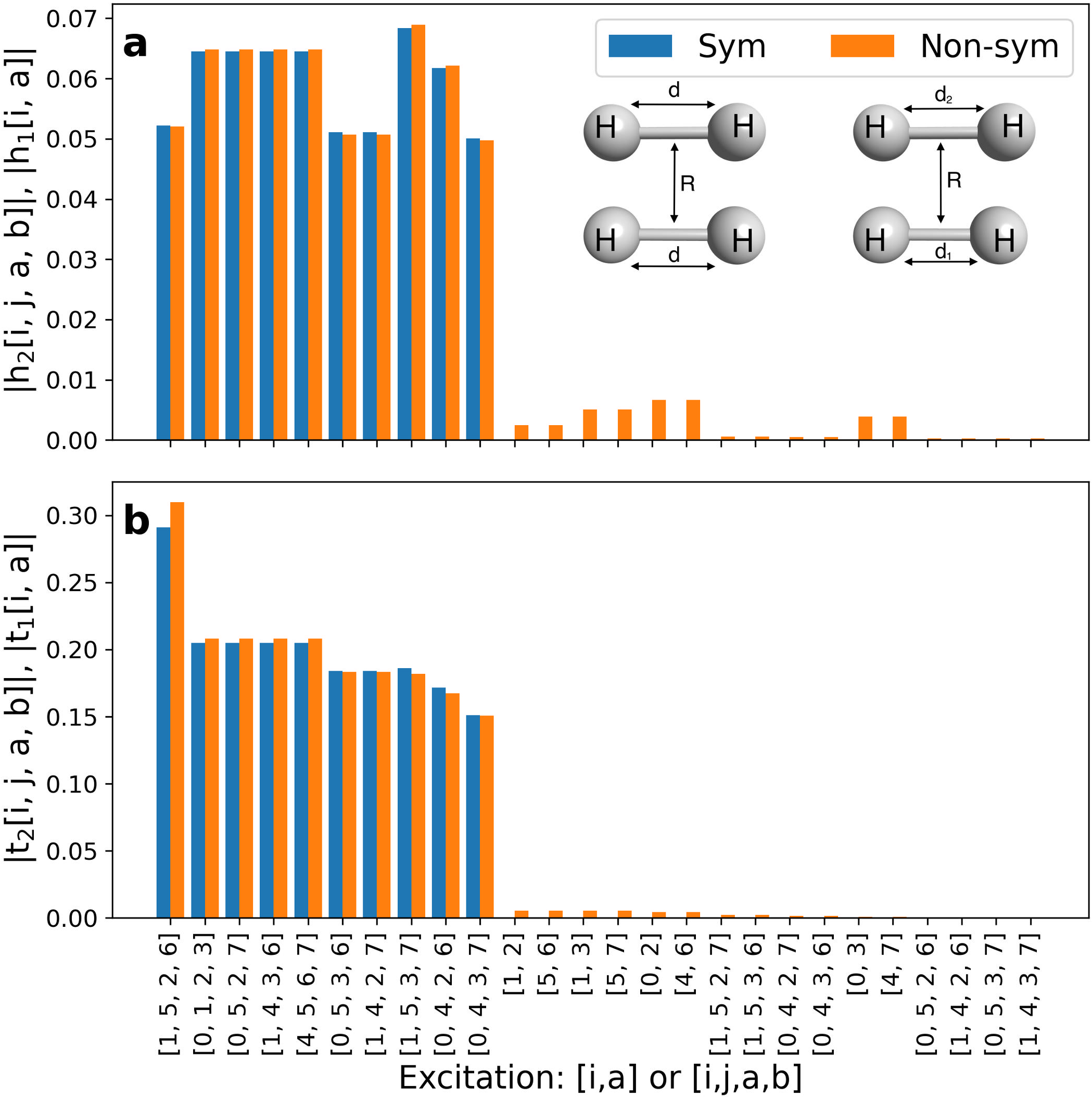}
    \caption{(a) Values of one- and two-electron integral matrix elements for the symmetric rectangular \ce{H4} molecule with $R = 1.5$ {\AA } and $d_1 = d_2 = 2.0$ {\AA } and the non-symmetric rectangular configuration with $R = 1.5$ {\AA }, $d_1 = 2.0$ {\AA }, $d_2 = 1.9$ {\AA }. (b) Values of UCCSD amplitudes $t_1[i, a]$, $t_2[i, j, a, b]$ for the symmetric rectangular \ce{H4} molecule with $R = 1.5$ {\AA }, $d_1 = d_2 = 2.0$ {\AA } and the non-symmetric square configuration with $R = 1.5$ {\AA }, $d_1 = 2.0$ {\AA }, $d_2 = 1.9$ {\AA }}
\end{figure}\label{fig:h4_rect}

After the first iteration, triple and quadruple excitations can be generated from the singles and doubles that are already included in the ansatz. There are multiple ways to obtain the coefficients in Equation \ref{uscc_selection} for these higher-order excitations. For example, by cycling through all single excitations $[i, a]$ present in the ansatz we can generate coefficients $t_1[i,a]h_2[j, k, b, c]$ with all possible doubles (already present in the ansatz or not). In addition, the coefficients $h_1[i,a]t_2[j, k, b, c]$ can be generated between singles and doubles present in the ansatz. Also, one can cycle through all doubles present in the ansatz to compute $t_2[i,j,a,b]h_1[k,c]$ coefficients, along with $h_2[i,j,a,b]t_1[k,c]$. All the coefficients mentioned above are associated with triple excitation $[i,j,k,a,b,c]$. Similarly, the coefficients describing quadruples $[i,j,k,l,a,b,c,d]$ include  $t_2[i,j,a,b]h_2[k,l,c,d]$ and all variations of indices involved. 
    
During each iteration $n$, starting from $n=2$, once all coefficients are computed for all excitations, the largest coefficient associated with an excitation is compared with the $\epsilon_n$ value corresponding to the current iteration. The $\epsilon_n$ for each iteration can be obtained from a predefined set of values or computed from the previous iteration, for example, $\epsilon_{n} = \epsilon_{n-1}/2$. A lower $\epsilon$ value results in more excitations being included in the ansatz, and the limit of $\epsilon=0$ corresponds to the full UCC ansatz, including all possible excitations up to desired order. Excitations for which the largest coefficient is larger than $\epsilon_n$ are added to the ansatz. In addition to triples and quadruples, it can include the singles and doubles that were not added during the first iteration. After the ansatz is constructed, the first VQE simulation is performed. If the set of excitations does not change from one iteration to another, the algorithm skips to the next $\epsilon$ value to avoid repeating the VQE energy evaluation. An alternative way to grow the ansatz is to arrange operators in decreasing order at each iteration so that a constant number of operators with the largest coefficients can be added at each iteration.  This iterative procedure is repeated until the termination condition. The latter can be reaching an $\epsilon$ value or a standard energy convergence criterion when the energy change between two iterations is smaller than the predefined $\epsilon_E$ value. It is important to point out that our selection criterion to choose excitation operators is a first-order crude approximation to ``importance'' criterion, and it will not provide the most compact wave function form like ADAPT-VQE. However, it does not require any additional calculations to compute the importance coefficients for an excitation operator. In addition, the inclusion of triple and quadruple excitations is straightforward. In this paper, we restrict ourselves to including only up to quadruple excitations; in general, this process can be used to select cluster operators of any order, which could be important for larger and more correlated systems.

\begin{algorithm}[H]
\caption{Unitary Selective coupled-cluster}
\begin{algorithmic}
\STATE \textbf{Step 1.} Generate single and double excitations for a given molecule.
\STATE ~~~~For all single and double excitations $[i,a]$ and $[i,j,a,b]$ in UCCSD add to ansatz all excitations for which $h_1[i,a]$ and $h_2[i,j,a,b]$ are larger than $\epsilon_1$.
\REPEAT
\STATE  \textbf{Step 2.} Run VQE with the current ansatz to compute energy, update amplitudes for each excitation present in ansatz.
\STATE  \textbf{Step 3.} For each single $[i, a]$ or double $[i, j, a, b]$ excitation present in ansatz using $t_1$ and $t_2$ values from the previous iteration and additional excitations $[k, c]$ or $[k, l, c, d]$ generate triple and quadruple excitations with the following coefficients:
\STATE ~~~$t_1[i,a]\cdot h_2[j, k, b, c]$
\STATE ~~~$h_1[i,a]\cdot t_2[j, k, b, c]$
\STATE ~~~$t_2[i,j,a,b]\cdot h_1[k,c]$
\STATE ~~~$h_2[i,j,a,b]\cdot t_1[k,c]$
\STATE ~~~$t_2[i,j,a,b]\cdot h_2[k,l,c,d]$
\STATE  \textbf{Step 4.} For each excitation, if the absolute value of the largest coefficient computed in step 3 is larger than $\epsilon_n$ on iteration $n$, add this excitation to ansatz.
\UNTIL{termination condition}
\end{algorithmic}
\end{algorithm}

\changed{
An important difference between the selected CI and selected coupled-cluster is that the coupled-cluster wave function truncated at certain order still contains so-called disconnected terms of higher order due to the nature of the wave function structure. For example, a determinant obtained by exciting 4 electrons from the Hartree-Fock configuration in Full CI has a coefficient $C_4$ that can be written in terms of coupled-cluster operators:

\begin{equation}\label{ci_vs_cc}
    C_4 = T_4 + T^2_2/2 + T_1T_3 + T_1^2T_2/2 + T_1^4/4!,
\end{equation}

\noindent where $T_4$ is a connected term, and the rest of the terms are disconnected. The power of the coupled-cluster wave function is that the contributions $T^2_2/2$, $T_1^2T_2/2$, $T_1^4/4!$ will be included even if CC wave function is truncated at the second-order (CCSD). $T_1T_3$ term will be included if we truncate at order 3. What our heuristic algorithm estimates are the importance coefficients for disconnected terms, where one of the amplitudes in the product is replaced by the electronic Hamiltonian matrix element, e.g. $h_1[i,a]t_2[j, k, b, c]$ instead of $t_1[i,a]t_2[j, k, b, c]$. Undoubtedly, there are systems for which the importance of disconnected terms does not correlate with the importance of the connected terms. However, we make this approximation in exchange for not having to run expensive calculations to estimate the importance of connected terms. A similar assumption has been adopted in the study by Lyakh and Bartlett, \cite{adaptive_cc} where the importance coefficients for higher-order excitations are based on the CI coefficients from CISD calculation that is run before the adaptive coupled-cluster calculation. More information regarding connected and disconnected coupled-cluster terms and connections to CI can be found in a comprehensive review by Bartlett and Musial\cite{cc_review_bartlett}
}

\section{Fermionic Operator Prescreening}\label{screening}

A well-known approach to prescreen the CC amplitudes is to use MP2 amplitudes. Romero et al.\cite{romero2018} performed a systematic study of the \ce{H4} molecule, where the energies were computed for different geometric configurations of the molecule using the VQE method and UCCSD ansatz with the excitation operators for UCCSD prescreened with different MP2 amplitude thresholds. The results showed that the number of operators in the ansatz was reduced by up to a factor of 3 for some geometries without sacrificing accuracy. It is important to point out that some of the single and double excitations in molecules can be screened out based on the symmetries present in a molecule. The fact that electronic Hamiltonian commutes with its symmetries $[H,S]=0$ has been routinely used in quantum chemistry to simplify calculations. It has also been applied in the context of CC theory.\cite{xu2018} However, in the context of UCC and quantum computing leveraging point group symmetries has not been as widespread. Cao et al.\cite{cao2021larger} recently published a study where they suggest a procedure for screening fermionic operators for UCCSD by directly applying point group symmetries. A similar technique has been applied to reduce the size of the operator pool in ADAPT-VQE ansatz.\cite{shkolnikov2021avoiding} \change{Alternatively, the method that uses mutual information can be applied to pre-screen the operators in a pool.\cite{Zhang_2021}}

 When the electronic Hamiltonian matrix elements connecting molecular orbitals are zero because of the symmetries, the UCC amplitudes connecting these orbitals are also zero because of the same symmetries. \change{Therefore, if we run a single iteration of USCCSD algorithm where single and double excitations are added to ansatz based on $h_1[i,a]$ and $h_2[i,j,a,b]$ values (see Algorithm 1) with a low threshold (i.e., $10^{-8}$), we obtain a UCCSD ansatz without the terms that are numerically zero due to symmetries. On its own this feature is not of great value because such terms can be removed by checking the symmetry. However, for slightly distorted geometries the more dominant excitations can be deduced from $h_1[i,a]$ and $h_2[i,j,a,b]$ terms, as demonstrated on the \ce{H4} system (Figure \ref{fig:h4_rect}), as discussed by Romero et al.\cite{romero2018}.} We consider the rectangular geometry with $R = 1.5$ {\AA }, $d_1 = d_2 = 2.0$ {\AA }. In the minimal STO-3G basis set, the \ce{H4} molecule has 26  excitations up to the second order: 8 singles and 18 doubles. In Figure \ref{fig:h4_rect}(a) we plot the values of electronic Hamiltonian matrix elements $h_1[i,a]$ and $h_2[i,j,a,b]$ connecting the occupied spin orbitals $i$, $j$ with virtual spin orbitals $a$, $b$  for $R_{H-H} = 1.5$ {\AA }. Figure \ref{fig:h4_rect}(b) depicts the $t_1$ and $t_2$ CCSD amplitudes connecting the same spin orbitals. 
For the symmetric configuration represented by the blue bars, \change{as expected,} the excitations for which electronic Hamiltonian matrix elements are zero also have zero UCCSD amplitudes. 
All of the $t_1$ UCCSD amplitudes are zero, and only 10 $t_2$ amplitudes are nonzero (see Figure ~\ref{fig:h4_rect}). The removal of these zero excitations does not affect the energy. In the broken symmetry configuration represented by orange bars in Figure \ref{fig:h4_rect} with $R_{H-H} = 1.5$ {\AA }, $d_1 = 2.0$ {\AA }, $d_2 = 1.9$ {\AA } many more excitations have to be included. Including only 10 excitations prescreened from the symmetric configuration results in additional error of 0.1 mHa above energy obtained with the full UCCSD ansatz. 
\change{Although such ansatz truncation results in accuracy loss, for the non-symmetric case depicted in Figure ~\ref{fig:h4_rect} sacrificing 0.1mHa of accuracy while reducing the number of excitation operators from 26 to 10 can be a good trade-off. Additional energy can be recovered by using a larger cut-off threshold. The contribution of these 16 terms will largely depend on geometry. We consider our zero-cost prescreening method advantageous compared to the MP2 prescreening, which has $O(N^5)$ scaling. In addition, this procedure can be applied in other methods, in which UCCSD fermionic operators are used to construct an ansatz, such as ADAPT-VQE, qubit-ADAPT-VQE, or adaptive-QITE (quantum imaginary time evolution).\cite{adaptive_qite}}

\section{Computational Details}
All calculations for this work were performed by using the Python code that we developed. All required electronic structure calculations on a classical computer, such as computation of one- and two-electron integrals, were performed in the PySCF package v. 1.7.6.\cite{Sun2018} The VQE simulations were performed by using the Qiskit-nature package v. 0.3.1.\cite{Qiskit} To map fermions to qubits, we used the Jordan-Wigner mapping.\cite{Jordan1928} The VQE simulations in Qiskit were performed by using the state vector simulator and SLSQP optimizer for finding the classical parameters. \change{For the first VQE-USCC iteration all parameters were initialized with zeros. Optimized values were used for consecutive iterations, which significantly accelerates the convergence. The number of USCC iterations required for VQE convergence ranged from a few to 20.} A minimal STO-3G basis set was used for all simulations unless explicitly stated otherwise. For all molecules, no molecular orbitals were frozen. 

\section{Unitary Selective coupled-cluster Simulations}

We tested our iterative unitary selective coupled-cluster method on the \ce{H6}, \ce{H2O}, and \ce{BeH2} set of molecules. Both symmetric and slightly distorted geometries were considered. To showcase the performance of the method in different regions of PESs, we chose three internuclear distances for each molecule: single, double, and triple equilibrium distances. All results were obtained by using cluster operator exponentiation. The USCCSDTQ energy convergence with respect to the number of parameters is depicted in Figure ~\ref{fig:USCCSDTQ6} with symmetric geometries in the top row and non-symmetric in the bottom. Each point corresponds to the parameter $\epsilon$ reduced by a factor of 2 from the previous iteration. \change{Data from simulations used to plot Figure ~\ref{fig:USCCSDTQ6} is provided in the Supplementary Information.}

\change{For systems where the UCCSD error is below the chemical accuracy threshold, our iterative method requires fewer parameters to achieve comparable accuracy (Fig. 2), up to a factor of 3 as in the case of \ce{H2O}.  For systems where the UCCSD error is larger than the chemical accuracy threshold, our method provides a way to recover extra correlation energy by adding triple and quadruple excitations. The addition of extra excitations according to our procedure gradually lowers the energy. The challenging case for our method is \ce{H_6} molecule. For the non-symmetric configuration, UCCSD achieves better accuracy than our method for $R$ values double and triple the equilibrium distance. This shows the limitations of our method, specifically the crude approximation for the screening coefficient computation, which can become unreliable in a system with a large number of excitations, each contributing little to correlation energy. This system demonstrates that it is much more challenging to study highly correlated systems when most of the possible electronic configurations have non-negligible contributions to the wave function. In symmetric \ce{BeH2} molecule, there is a region where the error is flat with the increasing number of operators added to ansatz. This means that some unimportant operators are being added to the ansatz ahead of the ones that reduce the energy in later iterations. This shows some limitations of our method. When the magnitudes of the amplitudes for different excitations are similar, which is the case for this stretched geometry, the coefficients we compute with the USCC method can become a poor approximation for choosing the important UCC amplitudes.}

\begin{figure*}
    \includegraphics[width=1.0\textwidth]{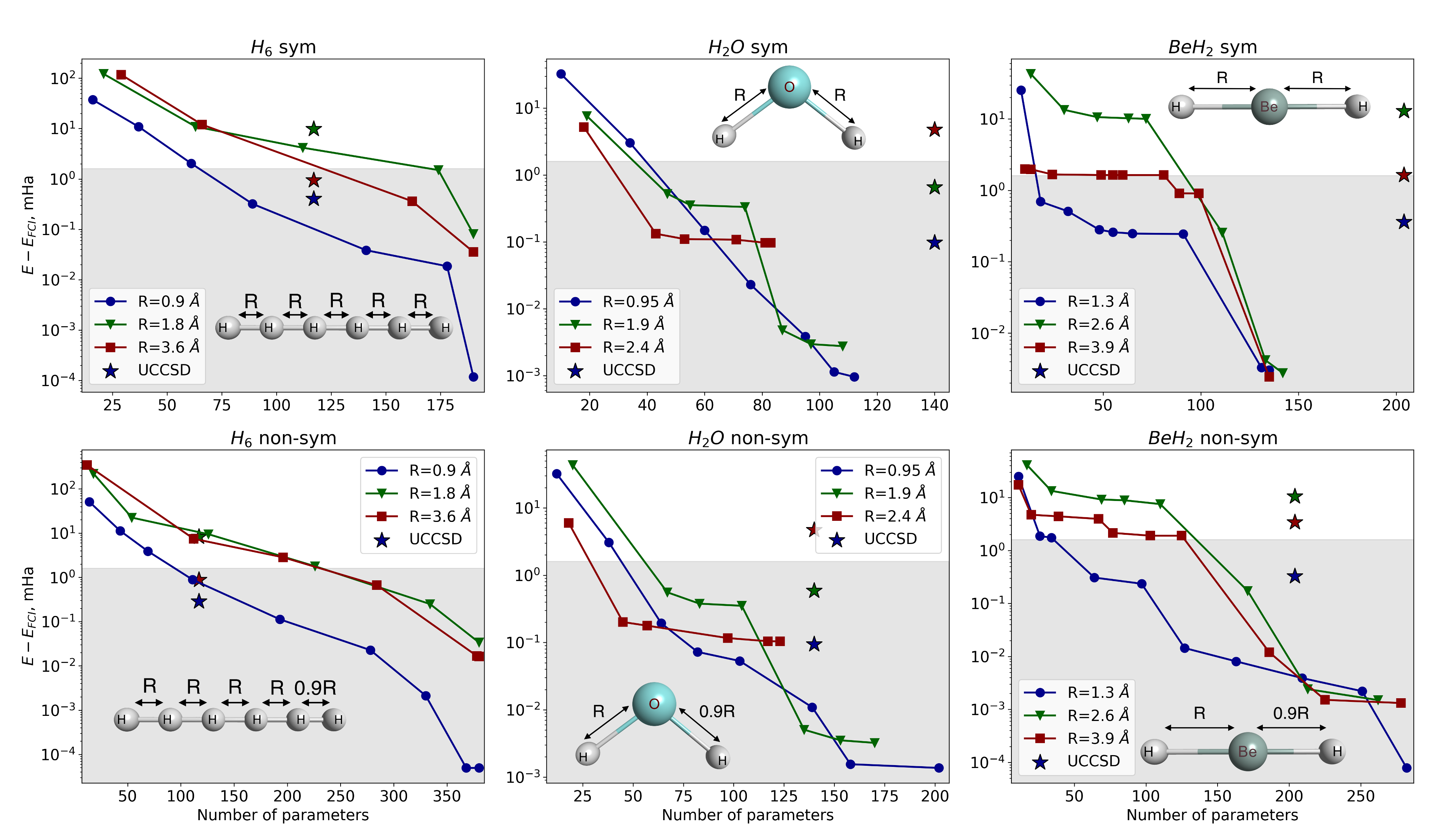}
    \caption{Error of the USCC simulation with respect to the number of parameters in the ansatz for symmetric (top row) and non-symmetric (bottom row) geometries of the \ce{H6}, \ce{H2O}, and \ce{BeH2} molecules. Colors represent the internuclear distances. UCCSD results are marked with stars. Each circle corresponds to a data point from each USCC iteration. Grey represents the area with errors smaller than chemical accuracy.}
\end{figure*}\label{fig:USCCSDTQ6}

The gate cost of implementing triple and quadruple excitations on quantum computers is high. \change{For example, for \ce{BeH2} molecule, the number of CNOT gates increases with the level of excitations: 12 for singles, 48 for doubles, 320 for triples, and 1792 for quadruples. Corresponding circuit depths are 21, 84, 464, and 2368 respectively.} Nevertheless, our proposed method offers a systematic and inexpensive way to expand the  UCC ansatz to improve accuracy. When comparing our USCC algorithm with adaptive methods such as ADAPT-VQE, we note that ADAPT-VQE will always construct a more compact ansatz because at every iteration it computes the energy gradients with respect to variational parameters for every element in the operator pool. The cost of such computations can be estimated as the measurement for up to $N^4$ Hamiltonian terms multiplied by the number of operators in the pool. The measurements of gradients for each variational parameter can be performed simultaneously because they are independent. For larger systems, however, the total cost in terms of shots can become infeasible. The USCC method has longer circuits but a much smaller shot count. A possible way to reduce the gate count is to use the qubit excitations instead of fermionic excitations.\cite{qubit_exc, iterative_qibit_exc} The cost of implementing qubit excitations is constant with respect to the number of qubits, which can dramatically reduce the gate count. However, ansatzes consisting of such excitations do not preserve the antisymmetry of the electronic wave function. Although such an approach has been shown to work for small molecules, studies for larger molecules are needed in order to definitively conclude that such an approach is scalable. \change{Another way to make the selection procedure more accurate is to improve the ''importance" criterion by using perturbation theory as it is done in selected CI. Removal of some high-order excitations in consecutive iterations if they do not contribute a significant amount of energy is also a possibility. This will be done in our future work.}

Most methods for VQE simulations that derive building blocks for ansatz construction from the fermionic excitations and do not go beyond double excitations because of the high cost. This  cost comes not just from the  large number of gates required to implement each triple or quadruple excitation but also because the total number of such operators also grows quickly. To compensate for the lack of higher-order excitations, one can use the UCCSD ansatz with generalized singles and doubles excitations (UCCGSD).\cite{uccgsd} In UCCGSD, excitations are generated between all orbitals, including virtual-virtual and occupied-occupied combinations. It allows one to recover more correlation energy than the standard UCCSD does. The number of such generalized excitations, however, grows much faster than standard UCCSD. To reduce the cost, the \textit{k}-UpCCGSD method uses only pair double excitations in the ansatz and increases the flexibility of the ansatz by using $k$ repetitions of the circuit. In the ADAPT-VQE method, only operators that contribute to correlation energy are added to the ansatz. The operator that will contribute the most to the correlation energy is picked based on the gradient of energy with respect to the variational parameter.\cite{Grimsley2019} The cost of computing gradients for such a large operator pool will result in a very high total shot count. Although it has been recently demonstrated that the size of the operator pool can be reduced by leveraging symmetries.\cite{shkolnikov2021avoiding} The same prescreening based on point group symmetries can be applied for constructing a more compact k-UpCCGSD ansatz without accuracy loss, or even k-UCCGSD, where all prescreened double excitations would be included, not just pair doubles.

In most studies describing new methods for constructing compact ansatzes for VQE simulations, the test molecules have a symmetric geometric configuration. We suggest that the performance of such truncated ansatzes should be tested against non-symmetric molecules as well, because the performance of such ansatzes might significantly depend on the geometry, unlike the complete ansatz, for example, UCCSD, which contains all possible excitations, and the performance difference between symmetric and non-symmetric systems is much less profound. We believe that prescreening based on symmetries should be always performed because it can significantly reduce the computational cost of most UCC-based existing methods with negligible overhead. 

\section{Conclusions}
We have proposed the selective coupled-cluster algorithm to construct unitary coupled-cluster ansatz with arbitrary order excitations. At each iteration, it uses the amplitudes of the excitations already included in ansatz and the values of the electronic Hamiltonian matrix elements to find important excitations of higher order to include in the ansatz. On a test set of small molecules, we have shown that it systematically improves accuracy at each iteration. However, it has limitations, for example at larger internuclear distances, where many electron configurations have contributions to the wave function similar in magnitude. \changed{These limitations may also be related to our selection criterion, which relies on the disconnected terms to estimate the importance of the connected terms in the coupled cluster expansion, as discussed in Section 3.}

In future work,  the method can be improved by utilizing a less crude approximation for the coefficients that define the importance of each excitation. This improvement can potentially be achieved by using perturbation theory when computing these ``importance coefficients," as is done in some flavors of selected CI. The current use of just electronic Hamiltonian matrix elements and cluster amplitudes from previous iterations serves as a proof of concept and is a  crude approximation to demonstrate the potential of such an approach. \changed{In the current form, the main advantage of our approach is the essentially zero computational cost associated with selecting new terms to add to the ansatz, which  requires just a look-up of electronic Hamiltonian matrix elements, the latter being precomputed before any quantum chemistry simulation, including VQE.}

\change{\section{Code Availability}
\change{The code to reproduce the data presented in the paper is publicly available in the form of GitHub repository: \hyperlink{https://github.com/dfedorov1988/USCC}{https://github.com/dfedorov1988/USCC}.}
}
\begin{acknowledgements}

This material is based upon work supported by Laboratory Directed Research and Development (LDRD) funding from Argonne National Laboratory, provided by the Director, Office of Science,  of the U.S. Department of Energy under Contract No. DE-AC02-06CH11357. It was also supported by the U.S. Department of Energy, Office of Science,
Office of Fusion Energy Sciences, under Award Number DE-SC0020249. Y.A.’s work at Argonne National Laboratory was supported by the U.S. Department of Energy, Office of Science, under contract DE-AC02-06CH11357.
This work was performed, in part, at the Center for Nanoscale Materials, a U.S. Department of Energy Office of Science User Facility, and supported by the U.S. Department of Energy, Office of Science, under Contract No. DE-AC02-06CH11357. Y.A.'s  work is also supported by the U.S. Department of Energy, Office of Science, National Quantum Information Science Research Centers. D.A.F's work is supported by General Atomics.

DISCLAIMER: This report was prepared as an account of work sponsored by an agency of the United States Government. Neither the United States Government nor any agency thereof, nor any of their employees, makes any warranty, express or implied, or assumes any legal liability or responsibility for the accuracy, completeness, or usefulness of any information, apparatus, product, or process disclosed, or represents that its use would not infringe privately owned rights. Reference herein to any specific commercial product, process, or service by trade name, trademark, manufacturer, or otherwise, does not necessarily constitute or imply its endorsement, recommendation, or favoring by the United States Government or any agency thereof. The views and opinions of authors expressed herein do not necessarily state or reflect those of the United States Government or any agency thereof.

\end{acknowledgements}

\end{document}